\documentclass[twocolumn,pra,showpacs,superscriptaddress]{revtex4-1}
\usepackage{graphicx}
\usepackage{subfigure}
\usepackage{amsmath}
\usepackage{bm}
\usepackage{float}
\usepackage{color}
\usepackage{overpic}
\usepackage{sidecap}
\allowdisplaybreaks
\usepackage{soul}
\setstcolor{red}
\usepackage{ulem}

\usepackage{hyperref}
\hypersetup{backref,pdfpagemode=FullScreen,colorlinks=true,breaklinks,urlcolor=blue,linkcolor=blue,citecolor=blue}

\begin{document}

\title{Transport through a quantum dot coupled to two Majorana bound states}

\author{Qi-Bo Zeng}
\affiliation{Department of Physics and State Key Laboratory of Low-Dimensional Quantum Physics, Tsinghua University, Beijing 100084, China}

\author{Shu Chen}
\affiliation{Beijing National Laboratory for Condensed Matter Physics, Institute of Physics,
Chinese Academy of Sciences, Beijing 100190, China}
\affiliation{Collaborative Innovation Center of Quantum Matter, Beijing, China}

\author{L. You}
\affiliation{Department of Physics and State Key Laboratory of Low-Dimensional Quantum Physics, Tsinghua University, Beijing 100084, China}
\affiliation{Collaborative Innovation Center of Quantum Matter, Beijing, China}

\author{Rong L\"u}
\email{rlu@mail.tsinghua.edu.cn}
\affiliation{Department of Physics and State Key Laboratory of Low-Dimensional Quantum Physics, Tsinghua University, Beijing 100084, China}
\affiliation{Collaborative Innovation Center of Quantum Matter, Beijing, China}

\begin{abstract}
We investigate electron transport inside a ring system composed of a quantum dot (QD) coupled to two Majorana bound states confined at the ends of a one-dimensional topological superconductor nanowire. By tuning the magnetic flux threading through the ring, the model system we consider can be switched into states with or without zero-energy modes when the nanowire is in its topological phase. We find that the Fano profile in the conductance spectrum due to the interference between bound and continuum states exhibits markedly different features for these two different situations, which consequently can be used to detect the Majorana zero-energy mode. Most interestingly, as a periodic function of magnetic flux, the conductance shows $2\pi$ periodicity when the two Majorana bound states are nonoverlapping (as in an infinitely long nanowire) but displays $4\pi$ periodicity when the overlapping becomes nonzero (as in a finite length nanowire). We map the model system into a QD--Kitaev ring in the Majorana fermion representation and affirm these different characteristics by checking the energy spectrum.
\end{abstract}

\pacs{73.21.-b, 74.78.Na, 73.63.-b, 03.67.Lx}

\maketitle
\date{today}

\section{Introduction}
Because of the growing interest in both fundamental physics and their potential applications
in topological quantum computation, Majorana fermions or Majorana bound states (MBSs)
have attracted much attention in recent years (for recent reviews, see Refs. \cite{SRElliott, TDStanescu, Alicea}).
After remaining elusive for about three quarters of a century, Majorana fermions, which are their own antiparticles
as originally discussed in the context of nuclear and particle physics,
are beginning to reveal their presence in solid state systems as emergent excitations
obeying non-Abelian statistics.
Many theoretical proposals have presented a variety of realizations for Majorana fermions,
including systems of one-dimensional (1D) p-wave superconductors or two-dimensional (2D) $p+ip$ superconductors
using topological insulators or conventional semiconductors \cite{MStone,PFendley,SRaghu,SBChung,RMLutchyn,YOreg},
ultracold atoms \cite{XJLiu,CQu,CChen,JRuhman}, superfluid He-3 \cite{NBKopnin,XLQi,SBChung2},
and chains of magnetic atoms on superconducting substrates \cite{SNadj,SNadj2,HYHui,EDumitrescu}, etc.
MBSs are predicted to exist at the ends of a semiconductor nanowire with strong spin-orbit coupling (SOC)
(for example in $InAs$ or $InSb$), external Zeeman field, and superconductivity induced by contacting with an s-wave superconductor \cite{RMLutchyn,TDStanescu2}.
A variety of detection schemes are experimentally implemented for their detection
and plausible signatures of MBSs, such as the conductance peak at zero bias \cite{VMourik,MTDeng,ADas,ADKFinck}
and the fractional Josephson effect \cite{DMBadiane,PAIoselevich,AMBlack}, have been reported,
yet, the "smoking gun" evidence for MBSs remains out of reach.

\begin{figure}[!ht]
\centering
\includegraphics[width=2.5in]{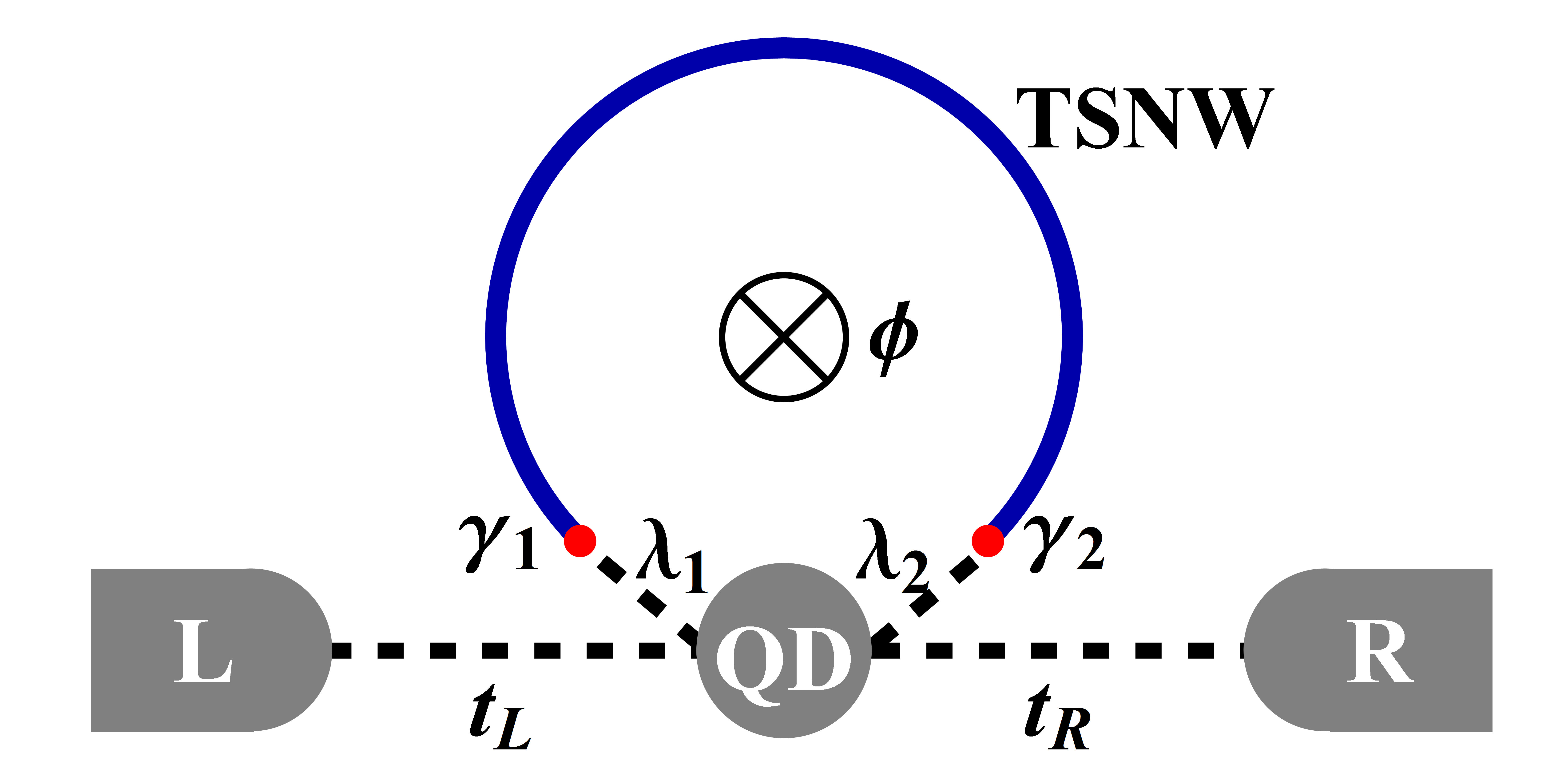}
\caption{(Color online) Schematic setup of the QD--MBSs ring system discussed in this paper. There are two Majorana bound states $\gamma_1$ and $\gamma_2$ confined at the two ends of a 1D topological superconductor nanowire (TSNW). A QD is coupled to these two MBSs; thus, the QD and the nanowire constitute a ring. Here, $\lambda_1= | \lambda_1 | e^{i\phi/4}$, $\lambda_2= | \lambda_1 | e^{-i\phi/4}$, $| \lambda_1 |$ and $| \lambda_2 |$ are the coupling strength, while $\phi$ is the phase resulting from the magnetic flux threading through the ring. Two normal metallic leads $L$ and $R$ are attached to the QD with coupling strength $t_L$ and $t_R$ so that the transport properties through the system can be detected.}
\label{fig1}
\end{figure}

Model systems based on hybrid structures are explored intensively as well
for the existence of Majorana fermions or MBSs, of which a very appealing type
involves a quantum dot (QD) coupled to MBSs.
These hybrid systems facilitate the study and the understanding of MBSs
 from their coupling to other systems.
Extensive efforts are directed at probing their transport properties such as
conductance spectrum, quantum interference,
and current noise for many different QD--MBS(s) configurations \cite{BHWu, AZazunov,AUeda,YCao,WJGong,HFLu1,HFLu2,HFLu3},
whose characteristic features are often used to implicate the existence of MBS.
Ref. \cite{KFlensberg} reports the non-Abelian operations on Majorana fermions
using Coulomb blockaded QDs. Recently, a ring system consisting of a topological
superconductor nanowire and normal metallic lead has been proposed to probe the quantum nonlocality of
Majorana fermions \cite{JDSau}. The conductance of this ring system shows
a $4\pi$-period as a function of the threaded magnetic flux. Ref. \cite{DELiu} predicts the conductance of a QD--MBS system
reduces to $e^2/2h$ in the topologically nontrivial state, compared to $e^2/h$ in the trivial state. An MBS--QD--MBS system is also investigated in \cite{DELiu}. However, the Fano effect and the change in periodicity in the conductance profile are not discussed, which will be covered and explained in detail in this paper.

In this work, we investigate electronic transport properties inside an analogous ring system
composed of a QD coupled to two MBSs located at the ends of a 1D topological superconductor nanowire (TSNW).
The ring geometry for our model system is shown in Fig. \ref{fig1}, whose
conductance through the QD is obtained with the help of the nonequilibrium Green's function method.
We find that it shows $2\pi$ periodicity when the wave function overlap between the two MBSs is zero
(as for a nanowire with infinite length).
For a nanowire with finite length where the MBSs display a nonzero overlap, the periodicity becomes $4\pi$.
The model system we consider is found to have zero-energy modes in the topologically nontrivial phase
only when the threading magnetic flux takes on the values of $\phi=\Phi/\Phi_0=(2n+1)\pi$.
Otherwise, there are no zero modes in this system.
Owing to quantum interference between bound and continuum states, the conductance spectrum for electrons
transporting through the system displays two Fano resonance peaks, which are symmetrically placed with respect to
the $eV=0.0$ line when $\phi=(2n+1)\pi$. When $\phi$ takes other values,
the corresponding Fano lineshape becomes asymmetrical.
We discuss our model system from both the Dirac fermion representation
as well as the Majorana fermion representation. A phenomenological explanation for the above-mentioned
characteristics is provided by mapping the ring system into a QD-Kitaev ring system
and with the different features verified by checking their corresponding energy spectrum.

This paper is organized as follows. In Sec. \ref{sec2},
the model Hamiltonian of a QD coupled to two MBSs is introduced and the conductance formula
calculated by using nonequilibrium Green's functions are presented.
Numerically computed results for transport properties and discussions on
the effects of the threaded magnetic flux are presented in Sec. \ref{sec3}.
Finally, a brief summary is provided in Sec. \ref{sec4}.

\section{Our model and the conductance formula}  \label{sec2}
As shown in Fig. \ref{fig1}, the ring model we consider in this paper
is composed of a QD coupled to two MBSs ($\gamma_1$ and $\gamma_2$) located at the ends
of a 1D TSNW. Through the two metal leads connected to the QD, we can detect
transport properties of the system, whose conductance spectrum can be tuned by changing the threading magnetic flux.
The Hamiltonian takes the following form
\begin{equation}
H = H_{\rm Leads} + H_{QD} + H_{MBS} + H_{DM} + H_{T},
\label{eqHam}
\end{equation}
where $H_{\rm Leads}$ in Eq. (\ref{eqHam}) represents the noninteracting electron gas in the left (L) and right (R) leads,
\begin{equation}
H_{\rm Leads} = \sum_{k,\alpha=L,R} \epsilon_{k\alpha} c_{k\alpha}^\dagger c_{k\alpha},
\end{equation}
where $c_{k\alpha}^\dagger$ and $c_{k\alpha}$ are the creation and annihilation operators with energy $\epsilon_{k\alpha}$
for the continuum in lead $\alpha$. $H_{QD}$ in Eq. (1) is the Hamiltonian of the quantum dot,
 \begin{equation}
H_{QD} = \epsilon_d d^\dagger d,
 \end{equation}
holding a single state at energy $\epsilon_d$, with $d^\dagger$ ($d$) being its creation (annihilation) operator.
The one-level model is assumed for simplicity. It can be easily extended to realistic cases involving more levels when needed.
The term in Eq. (\ref{eqHam})
\begin{equation}
H_{MBS} = i \epsilon_M \gamma_1 \gamma_2,
\label{mfs}
\end{equation}
describes the coupling between the two MBSs $\gamma_1$ and $\gamma_2$ with their overlap parameterized by $\epsilon_M$.
$H_{DM}$ denotes the linear coupling between the QD and the two MBSs according to
\begin{equation}
H_{DM} = (\lambda_1^* d^\dagger - \lambda_1 d) \gamma_1 + i (\lambda_2^* d^\dagger + \lambda_2 d) \gamma_2,
\end{equation}
with the coupling parameters $\lambda_1= | \lambda_1 | e^{i\phi/4}$, $\lambda_2= | \lambda_2 | e^{-i\phi/4}$,
where $| \lambda_1 |$ and $| \lambda_2 |$ denote the respective coupling strength,
and $\phi=\Phi/\Phi_0$ with $\Phi_0=h/2e$ is the phase factor resulting from the threading magnetic flux.
The last term in Eq. (\ref{eqHam})
\begin{equation}
H_{T} = \sum_{k\alpha} (t_\alpha c_{k\alpha}^\dagger d + h.c.),
\end{equation}
 represents the tunneling coupling between the QD and lead $\alpha$
with strength $t_\alpha$.

The two MBSs $\gamma_1$ and $\gamma_2$ in Eq. (\ref{mfs}) can be represented by their equivalent
Dirac fermion operators according to $\gamma_1=(f^\dagger + f)/ \sqrt{2}$ and $\gamma_2= i (f^\dagger - f)/ \sqrt{2}$,
which transforms the terms containing MBSs in the Hamiltonian Eq. (\ref{eqHam})
\begin{eqnarray}
H_{MBS} &=& \epsilon_M (f^\dagger f - \frac{1}{2}),\\
H_{DM} & =& \frac{1}{\sqrt{2}} (\lambda_1^* - \lambda_2^*) d^\dagger f^\dagger + \frac{1}{\sqrt{2}} (\lambda_1 - \lambda_2) f d \nonumber\\
&& + \frac{1}{\sqrt{2}} (\lambda_1^* + \lambda_2^*) d^\dagger f + \frac{1}{\sqrt{2}}(\lambda_1+\lambda_2) f^\dagger d.
\end{eqnarray}

The current from the lead $\alpha$ ($\alpha=L$ or $R$) to the central QD system is then given by
\begin{eqnarray}
I_\alpha &=& -e \langle \dot{N}_\alpha \rangle = \frac{ie}{\hbar}\langle [\sum_{k} c_{k\alpha}^\dagger c_{k\alpha}, H] \rangle \nonumber\\
&=& \frac{ie}{\hbar} \sum_{k} [t_\alpha \langle c_{k\alpha}^\dagger d \rangle - t_\alpha^* \langle d^\dagger c_{k\alpha} \rangle].
\end{eqnarray}
On introducing the relevant Green functions,
$G_{d, c_{k\alpha}^\dagger}^ < (t, t^\prime) = i \langle c_{k\alpha}^\dagger (t^\prime) d (t) \rangle$
and $G_{c_{k\alpha}, d^\dagger}^ < (t, t^\prime) = i \langle d ^\dagger (t^\prime) c_{k\alpha} (t) \rangle$, the above current becomes
\begin{eqnarray}
I_\alpha &=& \frac{e}{\hbar} \sum_{k} [ t_\alpha G_{d, c_{k\alpha}^\dagger}^ < (t, t) - t_\alpha^* G_{c_{k\alpha}, d^\dagger}^ < (t, t) ]\nonumber\\
& = & \frac{2e}{\hbar} \Re [ t_\alpha \sum_{k} G_{d, c_{k\alpha}^\dagger}^ < (t, t) ]\nonumber\\
& = & \frac{2e}{h} \Re \int d \omega t_\alpha \sum_{k} G_{d, c_{k\alpha}^\dagger}^ < (\omega).
\end{eqnarray}
Employing the method of nonequilibrium Green's functions, the above equation reduces to
\begin{equation}
I_\alpha = \frac{-2e}{h} \int d \omega \Gamma_\alpha \Im
\left[ \frac{1}{2} G_{d, d^\dagger}^ <  (\omega)  + f(\omega - \mu_\alpha)G_{d, d^\dagger}^ r (\omega) \right],
\end{equation}
where $\Gamma_\alpha = 2\pi i v | t_\alpha |^2 $ in the wide band approximation with $v$ being the density of states in the leads.
 $\Re(.)$ and $\Im(.)$ respectively denote the real and imaginary parts of $(.)$.
 $f(\omega - \mu_\alpha)$ is the Fermi-Dirac distribution with $\mu_\alpha$ the chemical potential for lead $\alpha$.
In the steady state, the current formula can be further simplified to
\begin{eqnarray}
I &= & \frac{-2e}{h} \times \nonumber\\
& &\int d \omega [f(\omega - \mu_L) - f(\omega - \mu_R)] \frac{\Gamma_L \Gamma_R}{\Gamma_L + \Gamma_R} \Im [G_{d, d^\dagger}^ r (\omega)], \nonumber
\end{eqnarray}
from which, the zero-temperature conductance is found to be
\begin{equation}
G = \frac{-2e^2}{h} \frac{\Gamma_L \Gamma_R}{\Gamma_L + \Gamma_R} \Im [G_{d, d^\dagger}^ r (\omega)] |_{\omega=eV},
\label{conductance}
\end{equation}
where we have set $\mu_L=eV$ and $\mu_R=0$. Thus the calculation of conductance requires the retarded dot Green function $G_{d, d^\dagger}^ r (\omega)$, which after some mathematical derivations is found to take the following form
\begin{equation}
G_{d, d^\dagger}^ r (\omega) = [\omega - \epsilon_d + \frac{i \Gamma}{2} - A(\omega) - B(\omega)]^{-1},
\label{dotgf}
\end{equation}
where
\begin{eqnarray*}
A(\omega) & =&  K(|\lambda_1|^2 + |\lambda_2|^2 + \frac{2\epsilon_M}{\omega} |\lambda_1| |\lambda_2| \cos \frac{\phi}{2}),\\
B(\omega) & =& \frac{K^2 (|\lambda_1|^4 + |\lambda_2|^4 - 2 |\lambda_1|^2 |\lambda_2|^2 \cos \phi)}{\omega + \epsilon_d + \frac{i\Gamma}{2} - K(|\lambda_1|^2 + |\lambda_2|^2 - \frac{2\epsilon_M}{\omega} |\lambda_1| |\lambda_2| \cos \frac{\phi}{2})},\\
\end{eqnarray*}
with $K$ and $\Gamma$ being defined as $K = \frac{\omega}{\omega^2 - \epsilon_M^2}$ and $\Gamma = \Gamma_L + \Gamma_R$.

Even without carrying out direct numerical computation, the above formula [Eqs. (\ref{conductance}) and (\ref{dotgf})] already show that
conductance is a periodic function
of magnetic flux $\phi$ with $2\pi$ period when the overlap of the two MBSs is zero,
i.e., when $\epsilon_M=0$. The period changes to $4\pi$ when $\epsilon_M$ is nonzero.
In the next section, we will present the detailed numerical results for the transport properties
of our model system containing a QD coupled to MBSs.

\section{Numerical results and discussions}  \label{sec3}
\begin{figure}[!ht]
\centering
\includegraphics[width=3.6in]{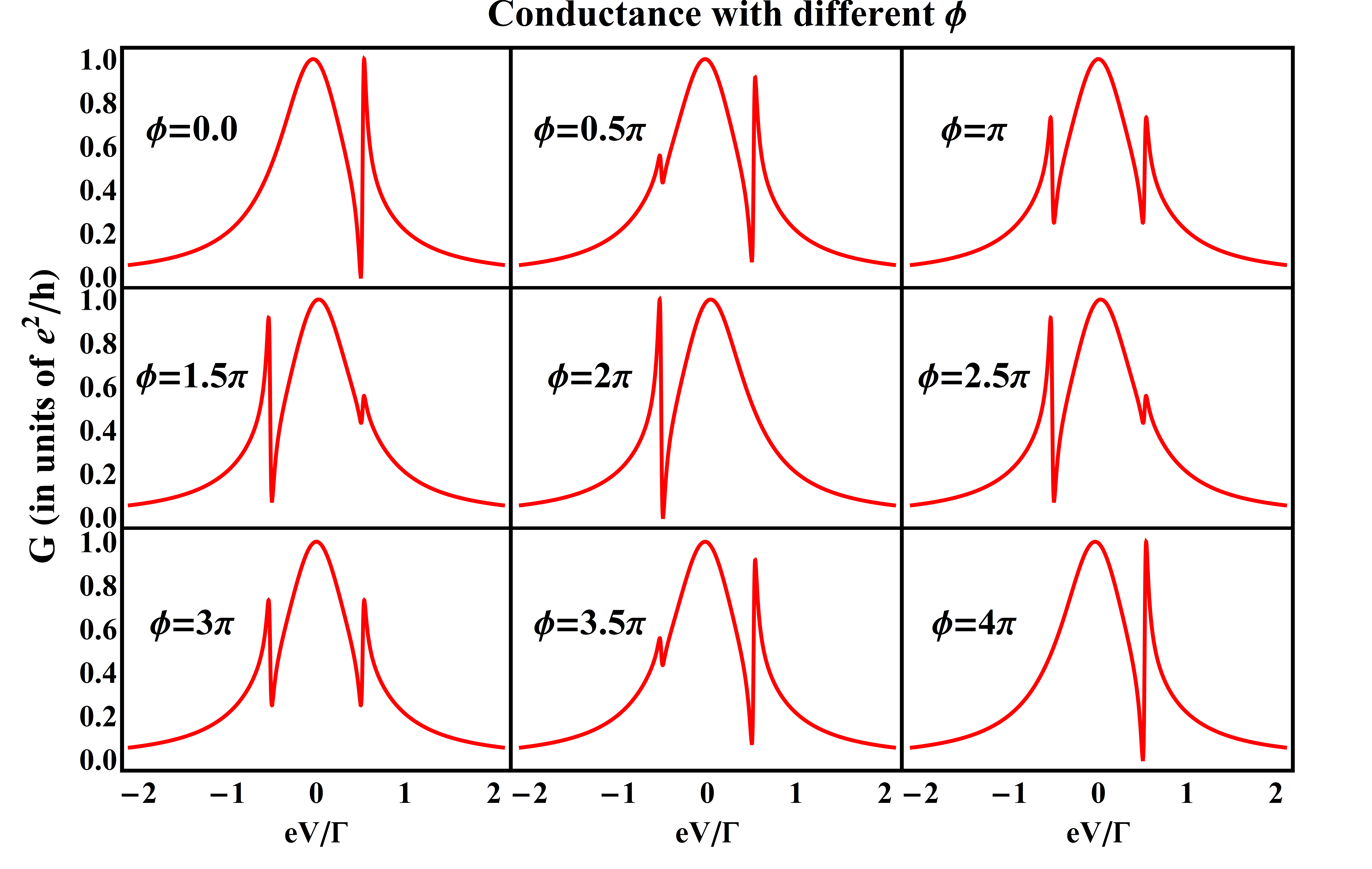}
\caption{(Color online) Conductance(in units of $e^2/h$) as a function of bias voltage $eV$ with different magnetic flux phase $\phi$ when $\epsilon_M=0.5\Gamma$. Other parameters: $|\lambda_1|=|\lambda_2|=0.1\Gamma$.}
\label{fig2}
\end{figure}

For simplicity, throughout this paper, we set $\epsilon_d=0$ and assume that the QD is symmetrically
coupled to the two MBSs, i.e., with $|\lambda_1| = |\lambda_2|$. Fig. \ref{fig2} shows the conductance
as a function of bias voltage $eV$. The dot-lead couplings is set to be symmetric $\Gamma_L=\Gamma_R$, and we take $\Gamma=\Gamma_L + \Gamma_R$ as the energy unit. The QD--MBSs couplings are set to be small $\lambda_1=\lambda_2=0.1\Gamma$ and
$\epsilon_M=0.5\Gamma$ for the overlap coupling between the two MBSs.
When the magnetic flux changes, conductance changes accordingly in
a periodic fashion with the period being $4\pi$.
The asymmetric Fano lineshape is clearly seen. More specifically, at $\phi=0.0$, $2\pi$, and $4\pi$,
only one Fano resonance shows up at around $eV=\epsilon_M$ or $-\epsilon_M$,
and the minimum conductance is suppressed to be zero. For other values of $\phi$, however,
two Fano resonances appear at around $eV=\pm \epsilon_M$ in the conductance spectra,
and the minima do not fall all the way to zero.
At odd integer multiples of the threaded magnetic flux $\phi=(2n+1)\pi$,
the two Fano resonances assume the same lineshape and are symmetrical about the $eV=0.0$ line. On the other hand, from the figure we can see that tuning the magnetic flux also leads to the swapping of the two resonance peaks.

The Fano resonances originate from the interference between electrons traversing in different paths when they travel from the left lead to the right lead. As the QD is strongly coupled to the leads, the phases of those electrons through the QD without going into the nanowire will not have distinct changes. However, those going into the nanowire will experience a swift phase changing (almost a $\pi$ phase changing). Thus the electrons through different paths will interfere with each other from destructively to constructively or vice versa over a small range of the parameter, which results in the asymmetric Fano lineshapes.

\begin{figure}[!ht]
\centering
\includegraphics[width=3.5in]{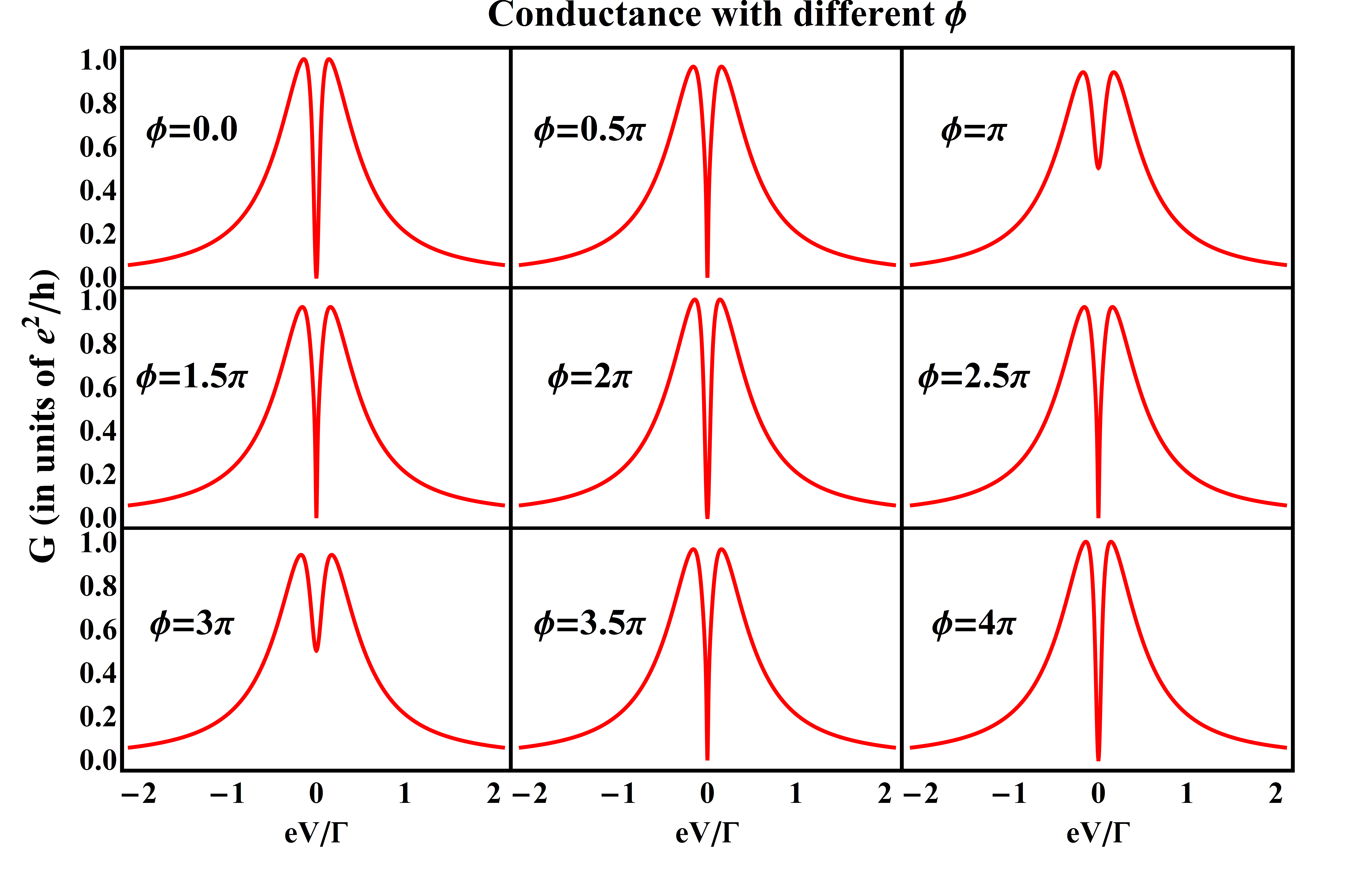}
\caption{(Color online) Conductance(in units of $e^2/h$) as a function of bias voltage $eV$ with different magnetic flux phase $\phi$ when $\epsilon_M=0.0$. Other parameters: $|\lambda_1|=|\lambda_2|=0.1\Gamma$.}
\label{fig3}
\end{figure}

\begin{figure}[!ht]
\centering
\includegraphics[width=3.0in]{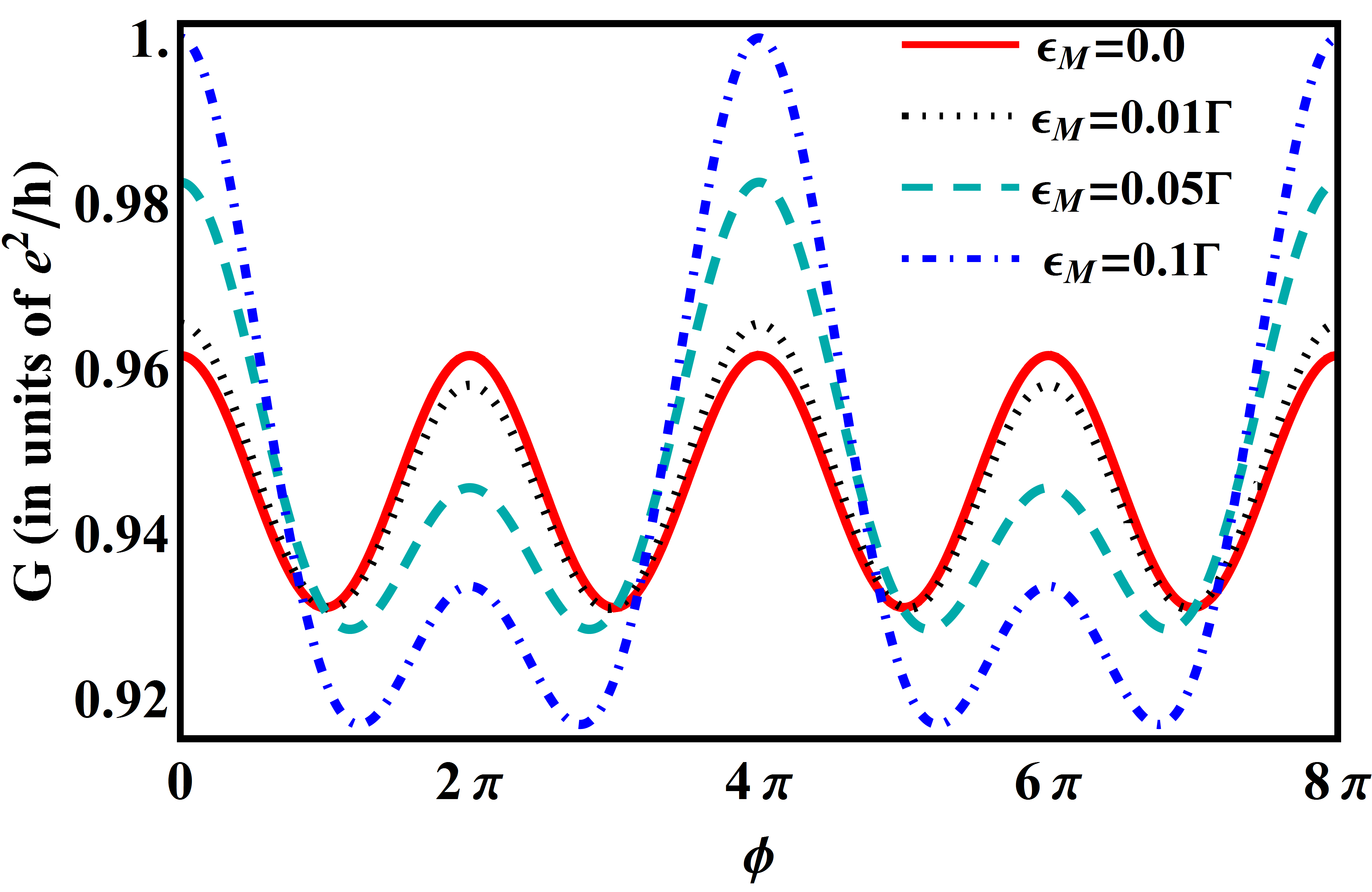}
\caption{(Color online) Conductance(in units of $e^2/h$) at $eV=0.2\Gamma$ as a function of the magnetic flux phase $\phi$ with different coupling strength between the two MBSs. When $\epsilon_M=0.0$, the period of conductance is $2\pi$; while for non-zero $\epsilon_M$, the period changes to $4\pi$. Other parameters: $|\lambda_1|=|\lambda_2|=0.1\Gamma$.}
\label{fig4}
\end{figure}

It is predicted \cite{DELiu} when the nanowire is in topological phase,
the conductance will fall to $e^2/2h$ at $eV=0.0$ if $\epsilon_M=0.0$ (namely when the two MBSs do not overlap).
The results from our model are presented in Fig. \ref{fig3}. It is interesting to note that at $\phi=\pi$ and $\phi=3\pi$,
conductance reduces to $e^2/2h$ at $eV=0.0$ while for other values of $\phi$, conductance is suppressed to zero at $eV=0.0$.
As $\epsilon_M$ becomes smaller and smaller, the two Fano resonances in Fig. \ref{fig2} get closer and closer
until they eventually merge into one resonance for $\epsilon_M=0$ at zero bias voltage. Most notably, the periodicity of conductance as a function of $\phi$ is $2\pi$, instead of $4\pi$ as shown in Fig. \ref{fig2}. In order to investigate the period change, we choose a specific value for $eV$, for example at $eV=0.2\Gamma$, and study the period of conductance under different choices of $\epsilon_M$, as shown in Fig. \ref{fig4}. The period remains $2\pi$ when the nanowire is infinitely long, namely, $\epsilon_M$ is zero. However, the period changes to $4\pi$ when $\epsilon_M$ deviates from zero for a nanowire with finite length.

In the following, by analyzing the Hamiltonian, we will discuss these different features of the system
in detail from both the Dirac fermion representation and the Majorana fermion representation.

\subsection{Dirac fermion representation}
\begin{figure}[!ht]
\centering
\begin{overpic}[width=3.3in]{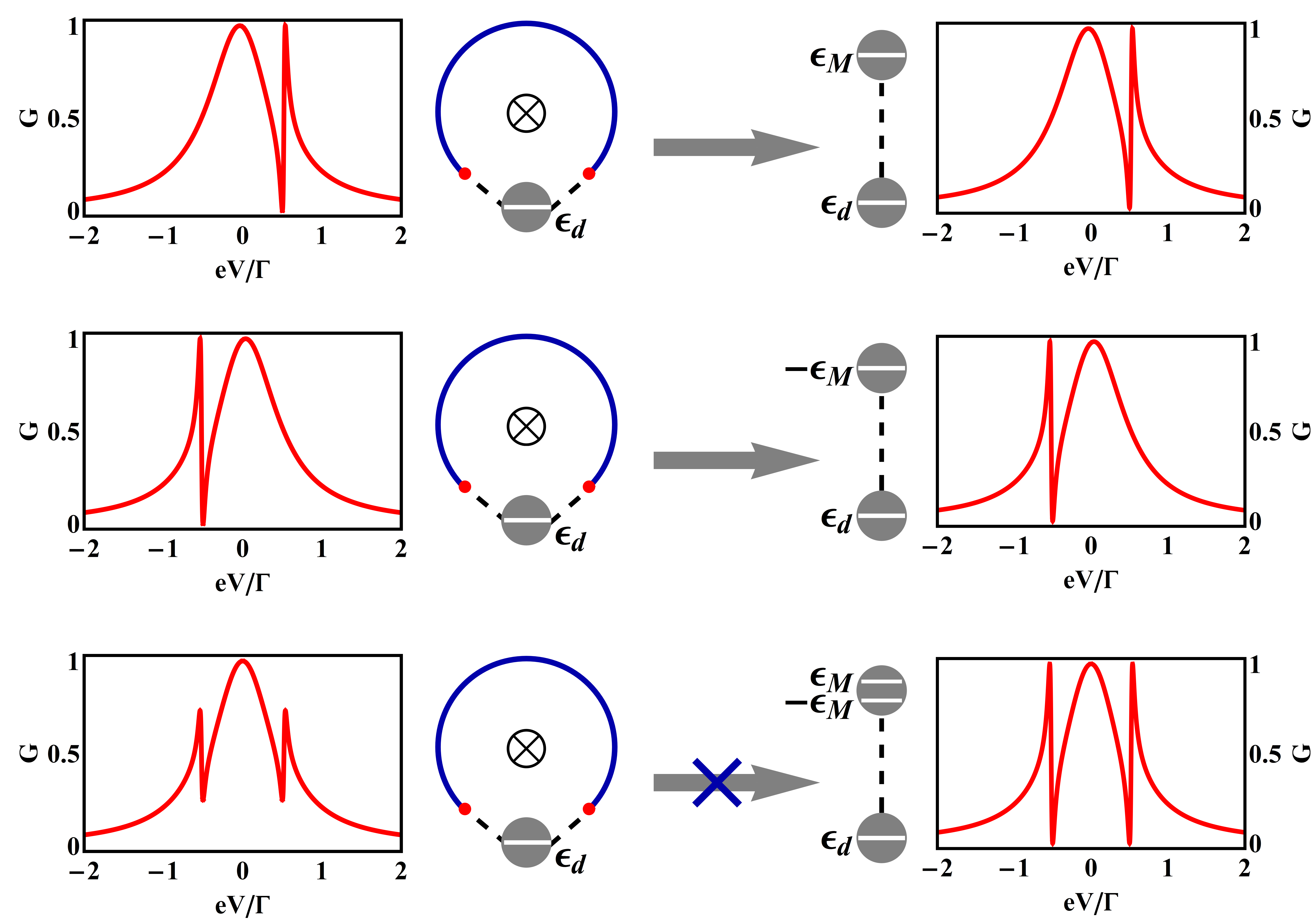}
\put(52,47){(a)}
\put(52,22){(b)}
\put(52,0){(c)}
\end{overpic}
\caption{(Color online) (a), (b) When $\phi$ takes an even integer multiple of $2\pi$, the QD--MBSs ring system can be transformed equivalently to a QD--QD model. (c) When $\phi=(2n+1)\pi$, the system cannot be transformed into a double-QD model. The parameters here are the same as in Fig. \ref{fig2}.}
\label{fig5}
\end{figure}

The MBSs $\gamma_1$ and $\gamma_2$ can be expressed in terms of the
regular Dirac fermions operators $\gamma_1=(f^\dagger + f)/ \sqrt{2}$ and $\gamma_2= i (f^\dagger - f)/ \sqrt{2}$.
If the QD is symmetrically coupled to the MBSs, our model system can be easily transformed
into a double-QD model under certain circumstances as shown in Fig. \ref{fig5}.

(i) \emph{Case $\phi=2n\times 2\pi$}. When $\phi$ takes an even integer multiple of $2\pi$,
the term describing the coupling between the QD and MBSs $H_{DM}$ can be simplified according to
\begin{equation}
H_{DM}=\sqrt{2}|\lambda_1|\cos(n\pi)d^\dagger f + \sqrt{2}|\lambda_1|\cos(n\pi)f^\dagger d ,
\end{equation}
in terms of the usual coupling between Dirac fermions.
The complete Hamiltonian thus is the same as the Hamiltonian
of a QD with energy level $\epsilon_d$ coupled to another QD with energy level $\epsilon_M$,
as shown in Fig. \ref{fig5}(a). The coupling strength between these two QDs is $\sqrt{2}|\lambda_1|\cos[n\pi]$.
Fig. \ref{fig5}(a) shows clearly that the conductance profiles for these two models
display the same behaviors when their respective parameters are the same.

(ii) \emph{Case $\phi=(2n+1)\times 2\pi$}. In this case, $H_{DM}$ reduces to
\begin{eqnarray}
H_{DM} &=& \sqrt{2}i |\lambda_1|\sin\left[(n+\frac{1}{2})\pi\right] f d \nonumber\\
&& - \sqrt{2}i |\lambda_1|\sin\left[(n+\frac{1}{2})\pi\right] d^\dagger f^\dagger,
\end{eqnarray}
where $f^\dagger$ and $f$ can be taken as the annihilation and creation operators
of the holes. We can make a transformation by setting $f^\dagger \rightarrow h$ and $f \rightarrow h^\dagger$,
according to which $H_{DM}$ becomes
\begin{eqnarray}
H_{DM} &=& \sqrt{2}i |\lambda_1|\sin\left[n+\frac{1}{2})\pi\right] h^\dagger d \nonumber\\
&&- \sqrt{2}i |\lambda_1|\sin\left[(n+\frac{1}{2})\pi\right] d^\dagger h.
\end{eqnarray}
The same transformation reduces the Hamiltonian MBSs $H_{MBS}=\epsilon_M(f^\dagger f - \frac{1}{2})$ to
$-\epsilon_M(h^\dagger h - \frac{1}{2})$. Putting all these terms together,
we observe that when $\phi=(2n+1)\times 2\pi$, the system is equivalent to a model in which
a QD with energy level $\epsilon_d$ is coupled to another QD with energy level $-\epsilon_M$,
and their coupling strength is $\sqrt{2}i |\lambda_1|\sin[(n+1/2)\pi]$ as shown in Fig. \ref{fig5}(b).
The conductance profiles of both models show Fano resonances at around $\omega=-\epsilon_M$.

(iii) \emph{Case $\phi=(2n+1)\pi$}. When $\phi=(2n+1)\pi$, we have
\begin{eqnarray}
&&H_{DM} = \nonumber \\
&& \sqrt{2}i |\lambda_1|\sin\left[\frac{(2n+1)\pi}{4}\right] f d - \sqrt{2}i |\lambda_1|\sin\left[\frac{(2n+1)\pi}{4}\right]  d^\dagger f^\dagger \nonumber\\
&&+\sqrt{2}|\lambda_1|\cos\left[\frac{(2n+1)\pi}{4}\right] d^\dagger f + \sqrt{2}|\lambda_1|\cos\left[\frac{(2n+1)\pi}{4}\right] f^\dagger d.\nonumber\\
\end{eqnarray}
By introducing the same operators for the holes as in the transformation above
for the case of $\phi=(2n+1)\times 2\pi$, the Hamiltonian $H_{DM}$ becomes
\begin{eqnarray}
&&H_{DM} = \nonumber\\
&& \sqrt{2}i |\lambda_1|\sin\left[\frac{(2n+1)\pi}{4}\right] h^\dagger d - \sqrt{2}i |\lambda_1|\sin\left[\frac{(2n+1)\pi}{4}\right] d^\dagger h\nonumber\\
&&+\sqrt{2}|\lambda_1|\cos\left[\frac{(2n+1)\pi}{4}\right] d^\dagger f + \sqrt{2}|\lambda_1|\cos\left[\frac{(2n+1)\pi}{4}\right] f^\dagger d,\nonumber\\
\end{eqnarray}
and $H_{MBS}$ reduces to $-\frac{1}{2}\epsilon_M h^\dagger h - \frac{1}{2}\epsilon_M + \frac{1}{2}\epsilon_M f^\dagger f + \frac{1}{2}\epsilon_M$
in this case, which resembles the system of a QD with two energy levels $\pm \epsilon_M$.
The complete system thus can be transformed into a model in which a QD with an energy level $\epsilon_d$
is coupled to another QD with two energy levels $\pm \epsilon_M$.
However, these two models are inequivalent, as shown from the conductance profiles in Fig. \ref{fig5}(c).
Although they show similar behaviors of two Fano resonances at around $\pm \epsilon_M$, the difference between these two models is significant. The conductance in the QD--MBSs system is not suppressed to zero at Fano resonance,
while the conductance in the QD--QD system is suppressed to zero.

The reason for the differences between these two models originates from the special
characteristics of the MBSs, which at the ends of the TSNW
can be represented by $\gamma_1=(f^\dagger + f)/\sqrt{2}$ and $\gamma_2=i (f^\dagger -f )/\sqrt{2}$.
The Majorana operators are self-adjoint, namely $\gamma_i^\dagger=\gamma_i$ $(i=1,2)$ and
thus they represent mixtures of particle and hole states.
If we replace $d^\dagger f^\dagger$ by $d^\dagger h$,
we will destroy the MBSs and render them into Dirac fermions. Thus, at $\phi=(2n+1)\pi$, the QD--MBSs ring system
is inequivalent to a double-QD system.
From the three typical cases considered above, we observe that when $\phi=2n\pi$, there will be no Majorana zero energy modes.
For $\phi=(2n+1)\pi$, on the other hand, zero energy modes will show up.
The electronic transport properties show different behaviors for these two different situations,
based on which one can distinguish the whether there are zero energy modes (or MBSs) in this system.

In addition, the mapping into the QD--QD model also allows for
an explanation of the different period for conductance
as a function of the threading magnetic flux $\phi$.
As shown above, when the overlap between the two MBSs changes from zero to finite values,
the period of conductance changes from $2\pi$ to $4\pi$.
According to the QD--QD models, the two cases of $\phi=2n\times 2\pi$ and $\phi=(2n+1)\times 2\pi$ cannot
be mapped to the same QD--QD model when $\epsilon_M$ is nonzero as the energy level of
the equivalent QD is $\epsilon_M$ for $\phi=2n\times 2\pi$ but is $-\epsilon_M$ for $\phi=(2n+1)\times 2\pi$.
Thus, the period of conductance is $4\pi$. However, for the special case of $\epsilon_M=0.0$,
the above two limits are mapped onto the same QD--QD models and the period of conductance becomes $2\pi$.

\subsection{Majorana fermion Representation}
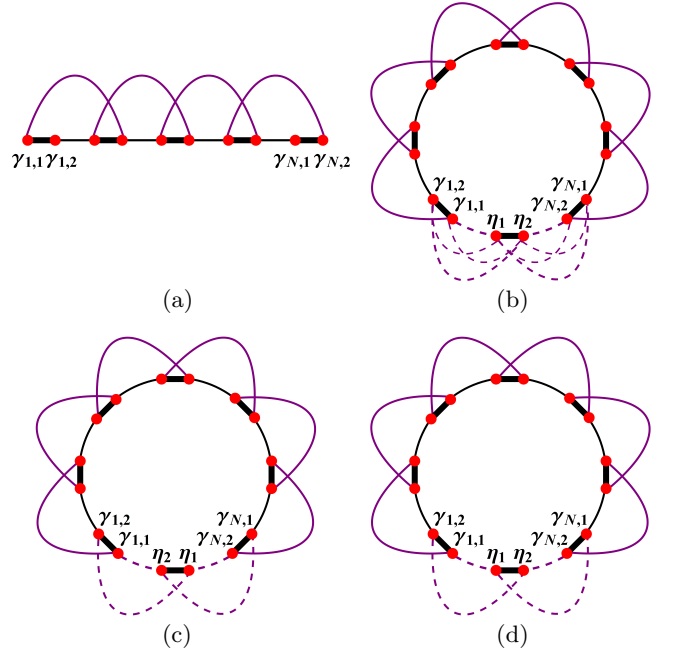
\begin{figure}[!ht]
\centering
\begin{overpic}[width=3.5in]{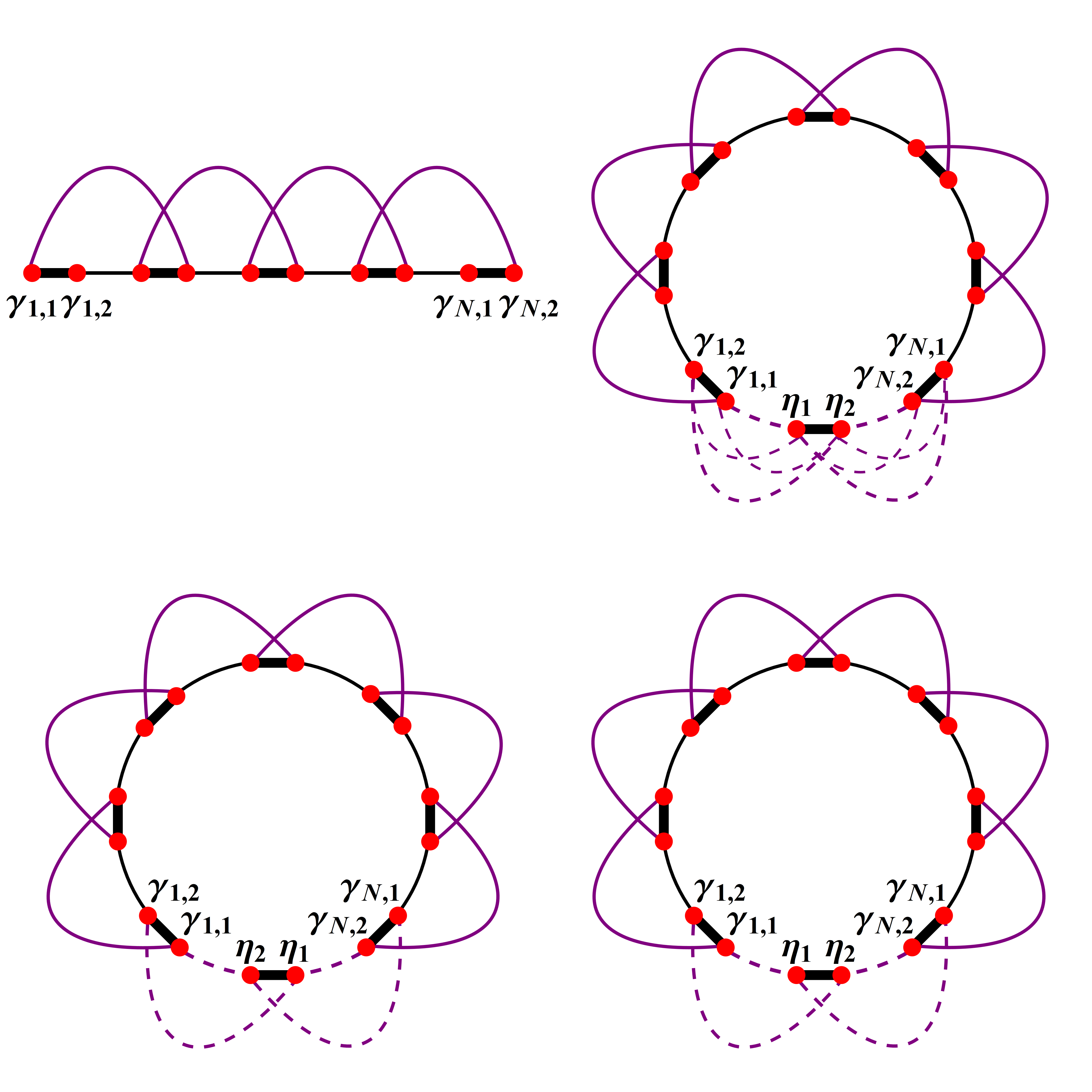}
\put(23,50){(a)}
\put(73,50){(b)}
\put(23,0){(c)}
\put(73,0){(d)}
\end{overpic}
\caption{(Color online) (a) Schematic illustration of a Kitaev chain. The red dot represents the Majorana bound states, the heavy black line indicates the coupling between $\gamma_{j,1}$ and $\gamma_{j,2}$, the light black line describes the coupling between $\gamma_{j,2}$ and $\gamma_{j+1,1}$ , and the solid purple line represents the coupling between $\gamma_{j,1}$ and $\gamma_{j+1,2}$; (b) The Kitaev chain and the MBSs $\eta_1$ and $\eta_2$ from the QD are coupled in a complicated way. The dashed purple line here shows the couplings between the MBSs from the QD and the MBSs from the chain; (c) When $\phi=2n\times 2\pi$, the Kitaev chain and the MBSs from the QD can form a Kitaev ring with the same coupling pattern as in the chain; (d) When $\phi=(2n+1)\times 2\pi$, the chain and the QD also form a Kitaev ring with the same coupling pattern in the chain.}
\label{fig6}
\end{figure}

Now we replace the superconductor nanowire by a Kitaev chain \cite{Kitaev},
which is composed of spinless fermions in a 1D lattice, as described below
\begin{eqnarray}
H_{K} &=& \sum_{j} \left[ -\mu(c_j^\dagger c_{j} - \frac{1}{2}) - t(c_j^\dagger c_{j+1} + h.c.)\right.\nonumber\\
&& \left.+ \Delta(c_{j+1}^\dagger c_{j}^\dagger + h.c.)\right],
\end{eqnarray}
with $t$ being the hopping amplitude between the lattice sites and $\Delta$ being the pairing amplitude.
The Hamiltonian for the QD and the coupling between the QD and the Kitaev chain are
\begin{eqnarray}
H_{QD} &=& \epsilon_d d^\dagger d,\\
H_{DK} &=& \lambda_1 c_1^\dagger d + \lambda_1^* d^\dagger c_1 + \lambda_2 c_N^\dagger d + \lambda_2^* d^\dagger c_N,
\end{eqnarray}
with $c_1$ and $c_N$ the operators for the first and the last lattice site, respectively.
Introducing Majorana operators $c_j=\frac{1}{\sqrt{2}}(\gamma_{j,1}+ i\gamma_{j,2})$ and $d=\frac{1}{\sqrt{2}}(\eta_{1}+ i\eta_{2})$,
we can express the Hamiltonian as
\begin{eqnarray}
H_K &=& i \sum_j \left[-\mu \gamma_{j,1} \gamma_{j,2} + (t+\Delta)\gamma_{j,2} \gamma_{j+1,1}\right.\nonumber\\
&& \left.+ (-t + \Delta) \gamma_{j,1} \gamma_{j+1,2}\right], \label{HKMF}\\
H_{QD} &=& i \epsilon_d \eta_1 \eta_2 + \epsilon_d, \label{HQDMF}\\
H_{DK} &=& -i|\lambda_1|\sin \left(\frac{\phi}{4}\right)  \eta_1 \gamma_{1,1} + i|\lambda_1|\cos \left(\frac{\phi}{4}\right)  \eta_1 \gamma_{1,2},\nonumber\\
&&-i |\lambda_1| \cos \left(\frac{\phi}{4}\right) \eta_2 \gamma_{1,1} -i |\lambda_1| \sin \left(\frac{\phi}{4}\right) \eta_2 \gamma_{1,2} \nonumber\\
&&+i|\lambda_2| \sin \left(\frac{\phi}{4}\right)  \eta_1 \gamma_{N,1} + i |\lambda_2| \cos \left(\frac{\phi}{4}\right)  \eta_1 \gamma_{N,2} \nonumber\\
&&-i|\lambda_2| \cos \left(\frac{\phi}{4}\right)  \eta_2 \gamma_{N,1} + i |\lambda_2| \sin \left(\frac{\phi}{4}\right) \eta_2 \gamma_{N,2}.\nonumber \label{HDKMF}\\
\end{eqnarray}
The redundancy of $\epsilon_d$ in $H_{QD}$ could be discarded. This allows for the analysis of the system
with different threading magnetic flux from the Majorana fermion perspective. Fig. \ref{fig6}(a) shows
the schematic setup of the Kitaev chain. When the chain is in the topologically trivial phase,
MBSs in the chain are all paired. If the chain is in the nontrivial phase,
unpaired MBSs will appear at the ends of the chain, namely $\gamma_{1,1}$ and $\gamma_{N,2}$
or $\gamma_{1,2}$ and $\gamma_{N,1}$, would become unpaired.
The complicated coupling between the QD and the Kitaev chain is shown in Fig. \ref{fig6}(b).

(i) \emph{Case $\phi=2n\times 2\pi$}. The Hamiltonian $H_{DK}$ could be written as
\begin{eqnarray}\label{casei}
H_{DK}&=&|\lambda_1|\cos(n\pi)(i\eta_1 \gamma_{1,2}) -|\lambda_1|\cos(n\pi)(i\eta_2 \gamma_{1,1}) \nonumber\\
 &&+ |\lambda_2| \cos(n\pi) (i\eta_1 \gamma_{N,2}) -|\lambda_2| \cos(n\pi) (i\eta_2 \gamma_{N,1}),\nonumber\\
\end{eqnarray}
in this case. We see that when $\phi=2n\times 2\pi$, the MBSs from the QD
are coupled to the chain in the same pattern as the MBSs on the chain.
Thus the QD and the Kitaev chain constitutes a Kitaev ring.
If $\gamma_{1,1}$ and $\gamma_{N,2}$ at the ends of the chain are unpaired, then from Eq. (\ref{casei}),
we observe that they could be paired with $\eta_2$ and $\eta_1$ from the QD, respectively.
If $\gamma_{1,2}$ and $\gamma_{N,1}$ are unpaired, they will instead be paired with $\eta_1$ and $\eta_2$, respectively.
So the whole QD-Kitaev chain system will have no unpaired MBSs or zero-energy modes even though the nanowire is in topologically nontrivial state, as shown in Fig. \ref{fig6}(c).

(ii) \emph{Case $\phi=(2n+1)\times 2\pi$}. In this case, the Hamiltonian $H_{DK}$ becomes
\begin{eqnarray}\label{caseii}
&& H_{DK} = \nonumber\\
&&-|\lambda_1| \sin\left(n+\frac{1}{2}\right)\pi (i\eta_1 \gamma_{1,1}) -|\lambda_1| \sin\left(n+\frac{1}{2}\right)\pi (i\eta_2 \gamma_{1,2})\nonumber\\
&& +|\lambda_2| \sin\left(n+\frac{1}{2}\right)\pi (i\eta_1 \gamma_{N,1})+|\lambda_2| \sin\left(n+\frac{1}{2}\right)\pi (i\eta_2 \gamma_{N,2}).\nonumber\\
\end{eqnarray}
Equation (\ref{caseii}) shows the same coupling pattern as that of the MBSs on the chain; thus,
the QD and the Kitaev chain also constitute a ring system. However, in this case $\gamma_{1,1}$ is paired with $\eta_1$ and $\gamma_{N,2}$ is paired with $\eta_2$
when $\gamma_{1,1}$  and $\gamma_{N,2}$ are unpaired in the chain. On the other hand, when $\gamma_{1,2}$ and $\gamma_{N,1}$ are unpaired, they will be coupled with $\eta_2$ and $\eta_1$, respectively, as shown in Fig. \ref{fig6}(d). Compared to the $\phi=2n\times 2\pi$ case, we observe that $\eta_1$ and $\eta_2$ are coupled with the different MBSs from the chain. The conductance profiles of these two situations are different as we discussed earlier (see Fig. \ref{fig5}(a) and (b)). This is the manifestation of the non-Abelian statistics abided by Majorana fermions.

(iii) \emph{Case $\phi=(2n+1)\pi$}. Substituting $\phi=(2n+1)\pi$ into $H_{DK}$, we can see that $\gamma_{1,1}$, $\gamma_{1,2}$, $\gamma_{N,1}$ and $\gamma_{N,2}$ are coupled to $\eta_1$ and $\eta_2$ with the same coupling strength, as shown in Fig. \ref{fig6}(b). The coupling pattern between the MBSs from the QD and from the chain is different from the pairing pattern for those MBSs inside the chain. The unpaired MBSs at the ends of the chain cannot be paired with the MBSs from the QD definitively; thus, the chain plus the QD does not constitute a Kitaev ring with all the MBSs paired up.
As a result, the QD--MBSs ring system has unpaired MBSs or zero-energy modes in the topologically nontrivial state and is distinct from the other two cases.

When $\phi$ takes other values apart from the three typical cases discussed above, the Hamiltonian $H_{DK}$
takes the same form as shown in Eq. (\ref{HDKMF}), except that for this general case
of $\phi$ the coupling amplitudes between MBSs are different.
For example, if $\sin \frac{\phi}{4} > \cos \frac{\phi}{4}$, the coupling between $\gamma_{1,1}$ ($\gamma_{1,2}$)
and $\eta_1$ ($\eta_2$) would be stronger, while the coupling between $\gamma_{N,2}$ ($\gamma_{N,1}$) and $\eta_1$ ($\eta_2$)
will be weaker. Thus, the unpaired MBSs at the ends of the Kitaev chain could always be paired up with one of the MBSs
from the QD through stronger coupling amplitudes. Thus, the Kitaev ring can be formed,
and all the MBSs would be paired up.
As a result, there are unpaired MBSs or zero-energy modes in this QD--MBSs ring system only when $\phi=(2n+1)\pi$.

\begin{figure}[!ht]
\centering
\subfigure[]{
\label{fig7(a)}
\includegraphics[width=1.5in]{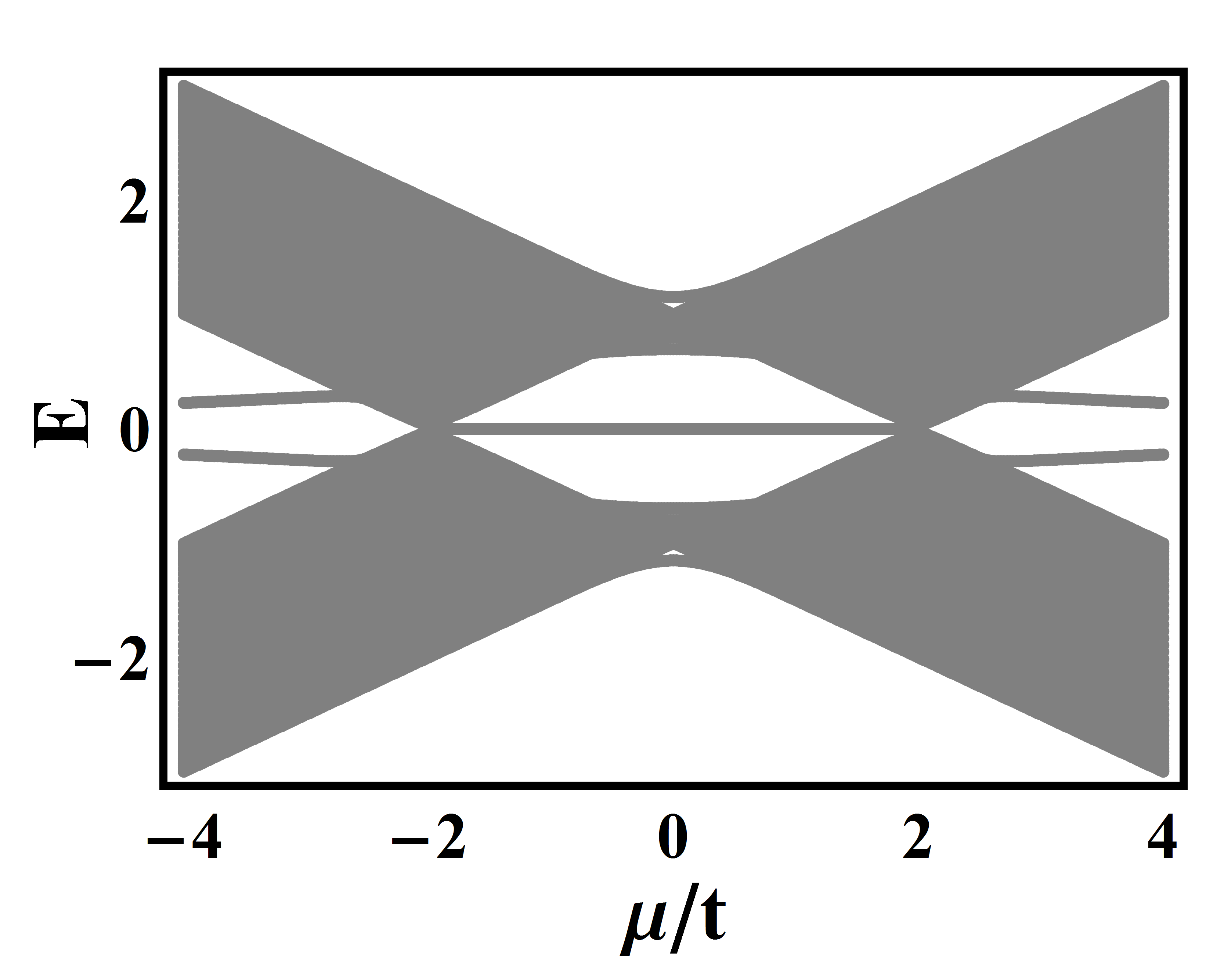}}
\subfigure[]{
\label{fig7(b)}
\includegraphics[width=1.5in]{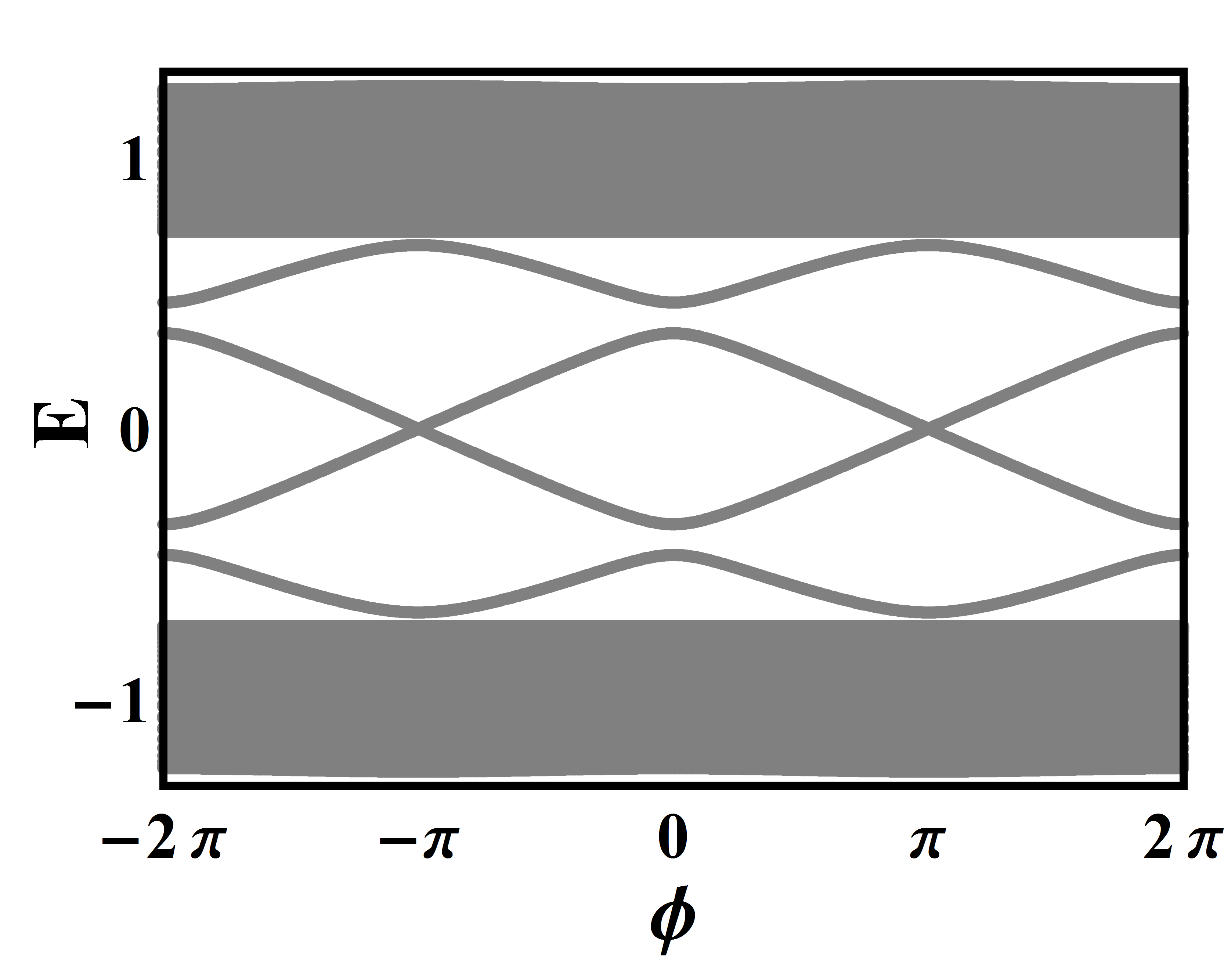}}
\subfigure[]{
\label{fig7(c)}
\includegraphics[width=1.5in]{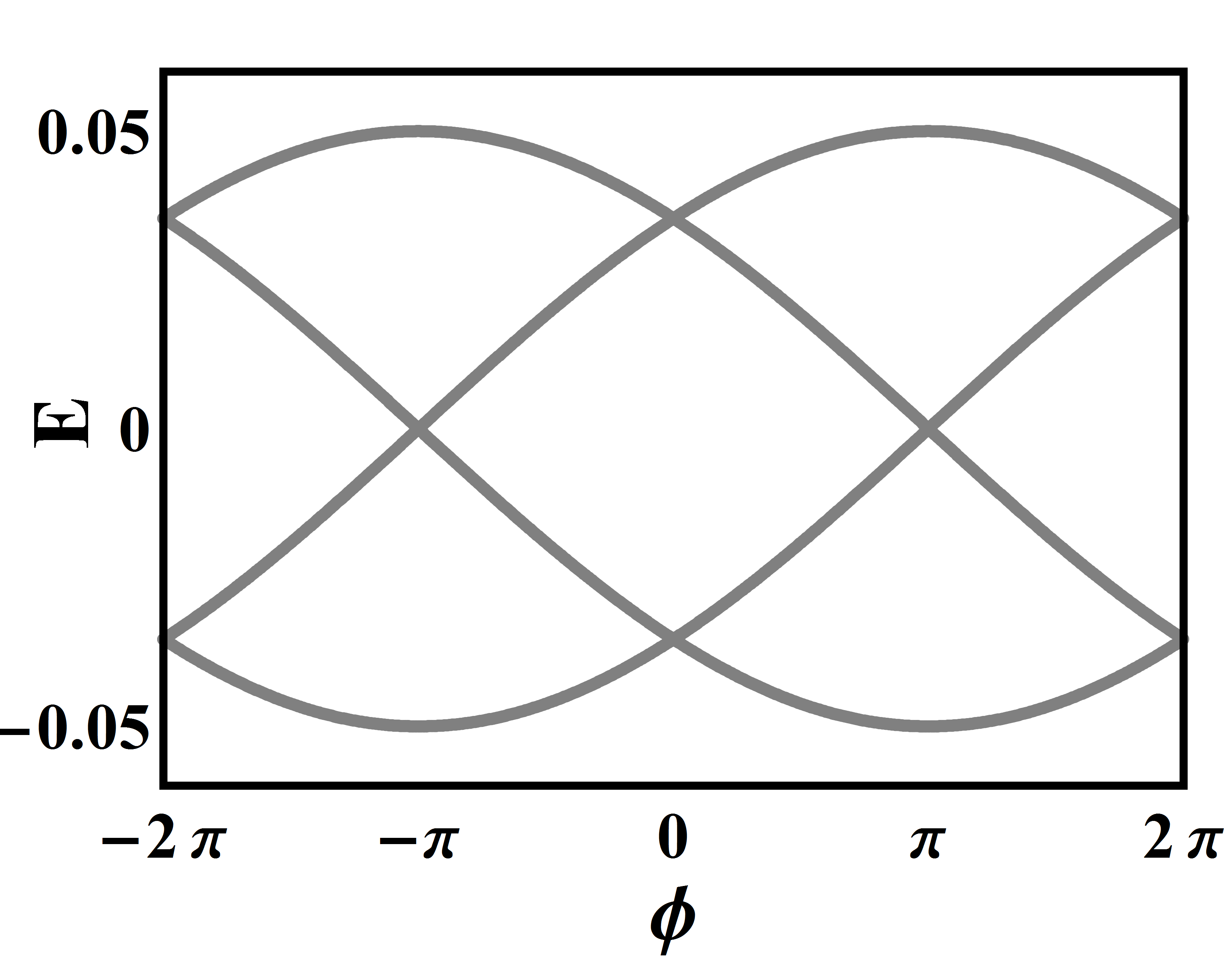}}
\subfigure[]{
\label{fig7(d)}
\includegraphics[width=1.5in]{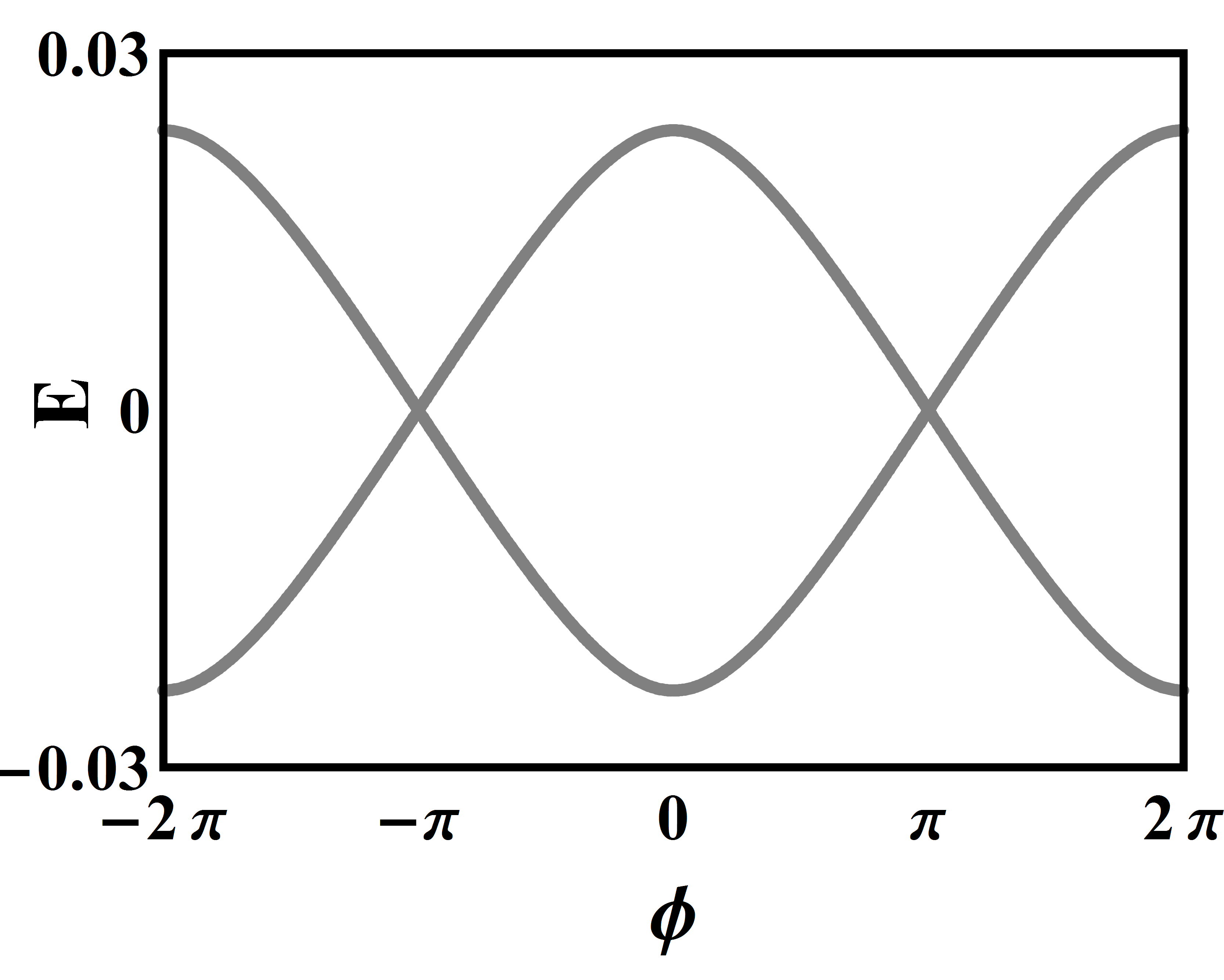}}
\caption{(a) The energy spectrum of the QD-Kitaev ring system as a function of $\mu$ when $\Delta=0.8$, $t=1$ and $\phi=\pi$. The other three pictures show the spectrum of the QD-Kitaev ring system as a function of magnetic flux $\phi$ with different parameters: (b) $\Delta=0.8$, $\mu=0.5$, $t=1$, $|\lambda_1|=|\lambda_2|=1$; (c) $\Delta=0.2$, $\mu=0$, $t=0.2$, $|\lambda_1|=|\lambda_2|=0.1$; (d) $\Delta=0.2$, $\mu=0.3$, $t=0.2$, $|\lambda_1|=|\lambda_2|=0.1$. The zero-energy modes only appear when $\phi=(2n+1)\pi$, otherwise the energy spectrum are always gapped.}
\label{fig7}
\end{figure}

We can further verify our above understanding by investigating the energy spectrum of the system.
In Fig. \ref{fig7(a)}, we show the energy spectrum as a function of $\mu$ for the QD-Kitaev ring system
with $\Delta=0.8$, $t=1$, and $\phi=\pi$. No zero mode in the gapped region is found
under the condition $|\mu| > 2t$, but zero modes show up when $\mu$ is smaller.
The Kitaev chain is in a topologically nontrivial state only under the condition $|\mu|< 2t$,
which is consistent with previous results \cite{SRElliott, KFlensberg}.
Figures \ref{fig7(b)}, \ref{fig7(c)}, and \ref{fig7(d)} show the energy spectrum of the QD-Kitaev ring system
as a function of the threaded magnetic flux for different parameters.
It shows that zero modes exist only under the condition $\phi=(2n+1)\pi$ when the Kitaev chain is in the nontrivial state, otherwise the energy spectrum will be gapped.

\begin{figure}[!ht]
\centering
\includegraphics[width=3.5in]{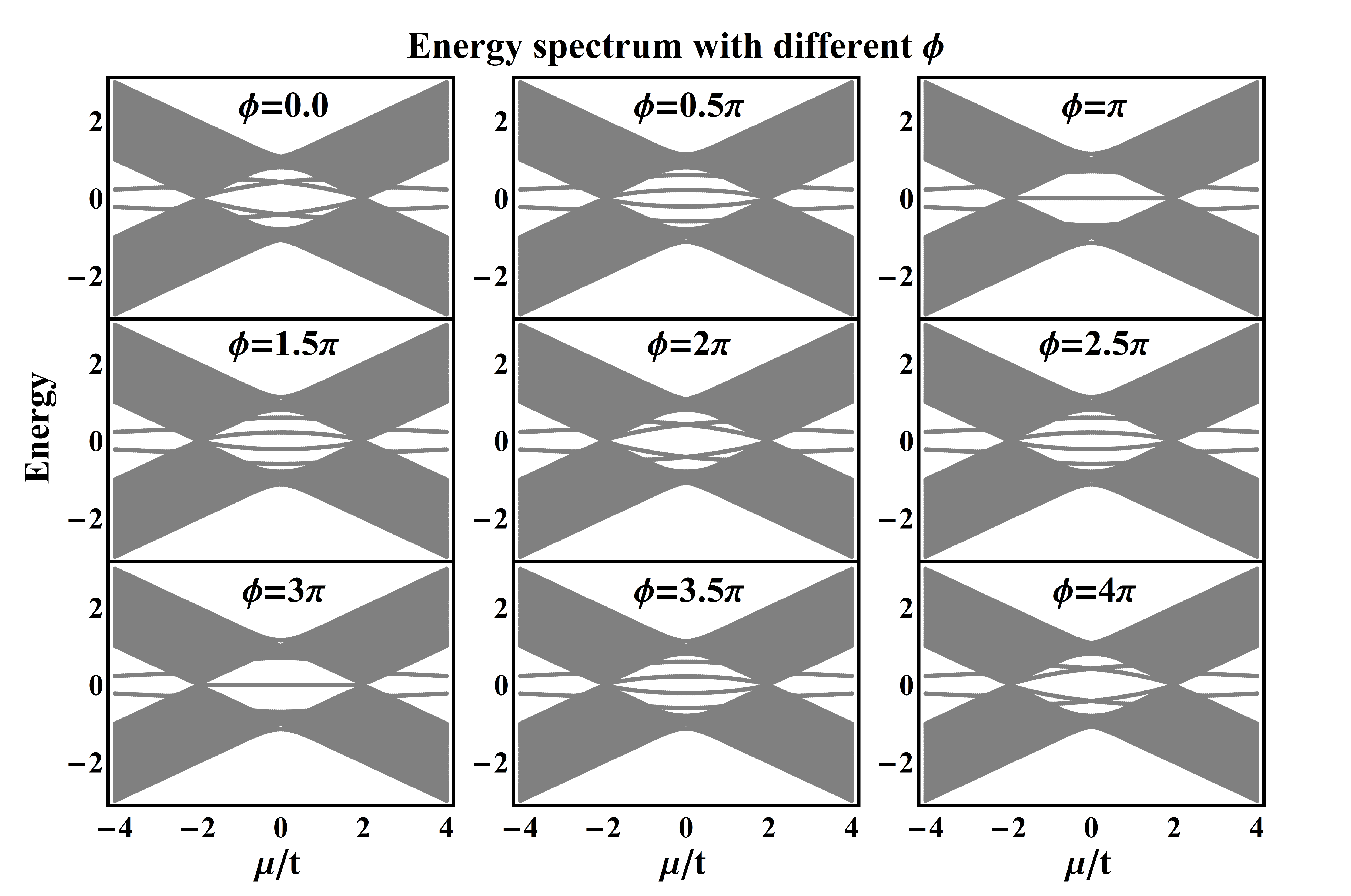}
\caption{Energy spectrum of the QD-Kitaev ring system. Here $\Delta=0.8$, $t=1$ and $|\lambda_1|=|\lambda_2|=1$. When $\phi=\pi$ and $3\pi$, the zero-energy modes appear only if $\mu<|2t|$. If $\phi$ takes other values, the spectrum is always gapped except at the points $|\mu|=2t$.}
\label{fig8}
\end{figure}

In Fig. \ref{fig8}, we show the energy spectrum of the QD-Kitaev ring system as a function of $\mu$ with
different threaded magnetic flux. The spectrum possesses the similar properties as the Kitaev chain.
When $\phi=(2n+1)\pi$ ($\phi=\pi$ and $3\pi$ in Fig. \ref{fig8}), zero-energy modes show up in the QD-Kitaev ring system when $|\mu|< 2t$, namely when the nanowire is in its topologically nontrivial phase. If the threaded magnetic flux takes other values, the energy spectrum is always gapped except at the zero-energy points $|\mu|=2t$, which are trivial as they are not robust against perturbations.

\section{Summary} \label{sec4}

We study the properties of electron transport through a QD coupled to two MBSs system in a ring configuration.
The MBSs at the two ends of a topological superconductor nanowire (TSNW) are strongly influenced by the threaded magnetic flux.
When $\phi=2n\times 2\pi$ (or $(2n+1)\times 2\pi$), the QD--MBSs ring system can be mapped into
an equivalent QD--QD system, in which the QD in the ring system is coupled to another QD
with energy levels $\epsilon_M$ (or $-\epsilon_M$), where $\epsilon_M$ denotes the overlap
between the two MBSs. When $\phi=(2n+1)\pi$, however, the system cannot be mapped into a QD--QD model.
The conductance spectra for these two situations are very different.
With appropriate parameters, the conductance shows one asymmetric Fano lineshape
at around $\omega=\epsilon_M$ (or $-\epsilon_M$) and the minimum of conductance
at the resonance is suppressed to zero when $\phi=2n\times 2\pi$ [or $(2n+1)\times 2\pi$].
When $\phi=(2n+1)\pi$, two Fano resonances appear with the same size at around $\omega=\pm \epsilon_M$
and the minima of the resonances do not decrease all the way to zero.
These distinctive features indicate whether there are Majorana zero-energy modes in the system.

It is also interesting to note that the conductance of the QD-two MBSs ring system shows
different periodicity for different coupling strength $\epsilon_M$ of the two MBSs. If the nanowire is infinitely long, i.e. $\epsilon_M=0.0$, the periodicity of conductance is $2\pi$. However, the periodicity changes
to $4\pi$ when $\epsilon_M$ deviates from zero as in a finite length nanowire.
We also investigate the transport properties of this QD-two MBSs ring system
 within the Majorana fermion representation by taking the electron on the QD
as the superposition of two MBSs and substituting the TSNW by a Kitaev chain.
We observe that when $\phi\neq(2n+1)\pi$, the unpaired MBSs from the Kitaev chain are separately
paired with one of the MBSs from the QD and thus constitutes a Kitaev ring in which all the MBSs are paired.
Thus, there is no unpaired MBSs for this case. However, when $\phi=(2n+1)\pi$,
the unpaired MBSs at the ends of the Kitaev chain cannot be paired up with MBSs from the QD definitely. Thus, there are unpaired MBSs or zero-energy modes in the the QD--MBSs ring system. The calculation of energy spectrum of the QD-Kitaev ring
system shows that zero-energy modes only exist under the condition $\phi=(2n+1)\pi$,
while for all other $\phi$ the system is always gapped except at
two nontopological zero points. The understanding gained from our work will be helpful for the ongoing experiments aimed at uncovering MBSs or Majorana zero mode through transport detections.

\section*{Acknowledgments}
This work has been supported by the NSFC under Grant No.11274195 and the National Basic Research Program of China (973 Program) Grant No. 2011CB606405 and No. 2013CB922000. Shu Chen is supported by National Program for Basic Research of MOST, by NSF of China under Grants No.11374354, No.11174360, and No.11121063, and by the Strategic Priority Research Program of the Chinese Academy of Sciences under Grant No. XDB07000000. L.Y. is supported by the MOST (Grant No. 2013CB922004) of the National Key Basic Research Program of China, and by NSFC (Nos. 91121005 and 91421305).


\begin{thebibliography}{99}

\bibitem{SRElliott}
S. R. Elliott and M. Franz, Rev. Mod. Phys. \textbf{87}, 137 (2015).
\bibitem{TDStanescu}
T. D. Stanesce and S. Tewari, J. Phys.: Condens. Matter \textbf{25}, 233201 (2013).
\bibitem{Alicea}
J. Alicea, Rep. Prog. Phys. \textbf{75}, 076501 (2012).
\bibitem{MStone}
M. Stone and R. Roy, Phys. Rev. B \textbf{69}, 184511 (2004).
\bibitem{PFendley}
P. Fendley, M. P. A. Fisher, and C. Nayak, Phys. Rev. B \textbf{75}, 045317 (2007).
\bibitem{SRaghu}
S. Raghu, A. Kapitulnik, and S. A. Kivelson, Phys. Rev. Lett. \textbf{105}, 136401 (2010).
\bibitem{SBChung}
S. B. Chung, H. J. Zhang, X. L. Qi, and S. C. Zhang, Phys. Rev. B \textbf{84}, 060510(R) (2011).
\bibitem{RMLutchyn}
R. M. Lutchyn, J. D. Sau, and S. Das Sarma, Phys. Rev. Lett. \textbf{105}, 077001 (2010).
\bibitem{YOreg}
Y. Oreg, G. Refael, and F. von Oppen, Phys. Rev. Lett. \textbf{105}, 177002 (2010).
\bibitem{XJLiu}
X. J. Liu, L. Jiang, H. Pu, and H. Hu, Phys. Rev. A \textbf{85}, 021603(R) (2012).
\bibitem{CQu}
C. Qu, Z. Zheng, M. Gong, Y. Xu, Li Mao, X. Zou, G. Guo, and C. Zhang, Nat. Commun. \textbf{4}, 2710 (2013).
\bibitem{CChen}
C. Chen, Phys. Rev. Lett. \textbf{111}, 235302 (2013).
\bibitem{JRuhman}
J. Ruhman, E. Berg, and E. Altman, Phys. Rev. Lett. \textbf{114}, 100401 (2015).
\bibitem{NBKopnin}
N. B. Kopnin and M. M. Salomaa, Phys. Rev. B \textbf{44}, 9667 (1991).
\bibitem{XLQi}
X. L. Qi, T. L. Hughes, S. Raghu, and S. C. Zhang, Phys. Rev. Lett. \textbf{102}, 187001 (2009).
\bibitem{SBChung2}
S. B. Chung and S. C. Zhang, Phys. Rev. Lett, \textbf{103}, 235301 (2009).
\bibitem{SNadj}
S. Nadj-Perge, I. K. Drozdov, B. A. Bernevig, and A. Yazdani, Phys. Rev. B \textbf{88}, 020407(R) (2013).
\bibitem{SNadj2}
S. Nadj-Perge, I. K. Drozdov, J. Li, H. Chen, S. Jeon, J. Seo, A. H. MacDonald, B. A. Bernevig, and A. Yazdani, Science \textbf{346}, 602 (2014).
\bibitem{HYHui}
H. Y. Hui, P. M. R. Brydon, J. D. Sau, S. Tewari, and S. Das Sarma, Sci. Rep. \textbf{5}, 8880 (2015).
\bibitem{EDumitrescu}
E. Dumitrescu, B. Roberts, S. Tewari, J. D. Sau, and S. Das Sarma, Phys. Rev. B \textbf{91}, 094505 (2015).
\bibitem{TDStanescu2}
T. D. Stanescu, R. M. Lutchyn, and S. Das Sarma, Phys. Rev. B \textbf{84}, 144522 (2011).
\bibitem{VMourik}
V. Mourik, K. Zuo, S. M. Frolov, S. R. Plissard, E. P. A. M. Bakkers, and L. P. Kouwenhoven, Science \textbf{336}, 1003 (2012).
\bibitem{MTDeng}
M. T. Deng, C. L. Yu, G. Y. Huang, M. Larsson, P. Caroff, and H. Q. Xu, Nano Lett. \textbf{12}, 6414 (2012).
\bibitem{ADas}
A. Das, Y. Ronen, Y. Most, Y. Oreg, M. Heiblum, and H. Shtrikman, Nat. Phys. \textbf{8}, 887 (2012).
\bibitem{ADKFinck}
A. D. K. Finck, D. J. Van Harlingen, P. K. Mohseni, K. Jung, and X. Li, Phys. Rev. Lett. \textbf{110}, 126406 (2013).
\bibitem{DMBadiane}
D. M. Badiane, M. Houzet, and J. S. Meyer, Phys. Rev. Lett. \textbf{107}, 177002 (2011).
\bibitem{PAIoselevich}
P. A. Ioselevich and M. V. Feigel'man, Phys. Rev. Lett. \textbf{106}, 077003 (2011).
\bibitem{AMBlack}
A. M. Black-Schaffer and J. Linder, Phys. Rev. B \textbf{84}, 180509 (2011).
\bibitem{BHWu}
B. H. Wu and J. C. Cao, Phys. Rev. B \textbf{85}, 085415 (2012).
\bibitem{AZazunov}
A. Zazunov, A. L. Yeyati, and R. Egger, Phys. Rev. B \textbf{84}, 165440 (2011).
\bibitem{AUeda}
A. Ueda and T. Yokoyama, Phys. Rev. B \textbf{90}, 081405(R) (2014).
\bibitem{YCao}
Y. Cao, P. Wang, G. Xiong, M. Gong, and X. Q. Li, Phys. Rev. B \textbf{86}, 115311 (2012).
\bibitem{WJGong}
W. J. Gong, Sh. F. Zhang, Zh. Ch. Li, G. Yi, and Y. S. Zheng, Phys. Rev. B \textbf{89}, 245413 (2014).
\bibitem{HFLu1}
H.-F. L\"u, H.-Z. Lu, and S.-Q. Shen, Phys. Rev. B \textbf{86}, 075318 (2012).
\bibitem{HFLu2}
H.-F. L\"u, H.-Z. Lu, and S.-Q. Shen, Phys. Rev. B \textbf{90}, 195404 (2014).
\bibitem{HFLu3}
H.-F. L\"u, H.-Z. Lu, and S.-Q. Shen. Shen, Phys. Rev. B \textbf{93}, 245418 (2016).
\bibitem{KFlensberg}
K. Flensberg, Phys. Rev. Lett. \textbf{106}, 090503 (2011).
\bibitem{JDSau}
J. D. Sau, B. Swingle, and S. Tewari, Phys. Rev. B \textbf{92}, 020511(R) (2015).
\bibitem{DELiu}
D. E. Liu and H. U. Baranger, Phys. Rev. B \textbf{84}, 201308(R) (2011).
\bibitem{Kitaev}
A. Y. Kitaev, Phys.-Usp. \textbf{44}, 131 (2001).

\end{thebibliography}
\end{document}